# Triiodothyronine suppresses humoral immunity but not T-cell-mediated immune response in incubating female eiders (*Somateria mollissima*).

Sophie Bourgeon and Thierry Raclot

*Institut Pluridisciplinaire Hubert Curien (IPHC), Département Ecologie, Physiologie et Ethologie (DEPE), UMR 7178 CNRS-ULP,*

*23 rue Becquerel, F-67087 Strasbourg Cedex 2, France*

Running title: Triiodothyronine and acquired immunity in eiders

Address for correspondence:

Sophie Bourgeon

IPHC, Département Ecologie, Physiologie et Ethologie, UMR 7178 CNRS-ULP,

23 rue Becquerel, F-67087 Strasbourg Cedex 2, France

Tel: (33) 3.88.10.69.15

Fax: (33) 3.88.10.69.06

e-mail:sophie.bourgeon@c-strasbourg.fr

**ABSTRACT**

Immunity is believed to share limited resources with other physiological functions and this may partly account for the fitness costs of reproduction. Previous studies have shown that the acquired immunity of female common eiders (*Somateria mollissima*) is suppressed during the incubation fast. To save energy, triiodothyronine ($T_3$) is adaptively decreased during fasting in most bird species, despite $T_3$ levels are maintained throughout incubation in female eiders. However, the relationship between thyroid hormones and the immune system is not fully understood. The current study aimed to determine the endocrine mechanisms that underlie immunosuppression in incubating female eiders. To this end we assessed the effects of exogenous $T_3$ on both components of the acquired immune system in 42 free-ranging incubating birds. Half of the females were implanted with $T_3$ pellets, while the other half sham implanted served as control. We measured variations in the immunoglobulin index, the T-cell-mediated immune response, body mass, and plasma parameters in both groups before and after manipulation. $T_3$ levels in implanted females were 4 times higher and mass loss was 40 % greater than in control females. Implanted females also showed an 18 % decrease in the immunoglobulin index, while the T-cell-mediated immune response was not significantly affected by the treatment. Finally, the treatment did not induce any significant changes in corticosterone levels. Our study shows that exogenous $T_3$ decreased only one component of the acquired immune system. We suggest that the immunosuppressive effect of $T_3$ could be mediated by its effects on body fat reserves. Further experiments are required to determine 1) the relationship between adiposity and immune function, 2) the adaptive significance of immunosuppression during incubation in eiders.

**INTRODUCTION**

Life-history theory predicts a trade-off between an organism's current reproductive effort and its future survival and reproductive success, where an increased reproductive effort could be deleterious to adult survival (Williams, 1966; Stearns, 1992). One potential mechanism whereby current reproductive effort may incur long-term reproductive costs is through a resource shift away from the immune system (Gustafsson et al., 1994; Sheldon and Verhulst, 1996; Råberg et al., 1998). Acquired immunity consists of two components (Roitt et al., 1998): humoral immunity (mediated by B-lymphocytes) and cell-mediated immunity (mediated by T-lymphocytes). Previous studies have shown that the acquired immunity of female common eiders (*Somateria mollissima*) is suppressed during the incubation fast (Hanssen et al., 2004; Bourgeon et al., 2006a) while its experimental activation has strong negative effects on the fitness of female eiders (Hanssen et al., 2004; Hanssen et al., 2005; Hanssen, 2006). However, Bourgeon et al. (2006a) found that both components of the acquired immunity decrease independently of each other. Whatever the adaptive significance of such an immunosuppression in breeding eider ducks may be, the underlying physiological mechanisms are also largely unknown. Deerenberg et al. (1997) suggested that a reduced immunocompetence during reproduction could be related to hormonal changes.

Thermoregulation and energy metabolism are partially regulated by triiodothyronine ($T_3$) in mammals and birds (McNabb, 1995). While female eider ducks fast throughout incubation (Korschgen, 1977; Gabrielsen et al., 1991; Criscuolo et al., 2000), they also have to maintain a high body temperature, required for egg incubation (Criscuolo et al., 2003). Criscuolo et al. (2003) showed that $T_3$ levels in female eiders were maintained throughout incubation. Circulating $T_3$, thermogenesis, and basal metabolic rate (BMR) were found to be positively correlated in precocial species during fasting (Sechman et al., 1989; Gabarrou et

1  al., 1997). Furthermore, daily energy expenditure and the T-cell-mediated immune response
2  were negatively correlated in female pied flycatchers (*Ficedula hypoleuca*; Moreno et al.,
3  2001). However, a depressed level of thyroid hormones led to a suppressed cytotoxic T-cell
4  activity in mallard ducks (*Anas platyrhynchos*) (Fowles et al., 1997), while it increased the *in*
5  *vitro* T-cell proliferative response to mitogenic stimulation in chickens (Williamson et al.,
6  1990). Hence, the relationship between thyroid hormones and the immune system is far from
7  understood (Smits et al., 2002) and remains to be clarified.
8        The main objective of our study was to examine the effects of increased plasma $T_3$
9  levels on both components of the acquired immunity in free-ranging female common eiders
10  during their incubation phase. To this end, incubating females were implanted with $T_3$ pellets
11  and compared to control birds. We subsequently measured variations of the female
12  immunoglobulin index, the T-cell-mediated immune response, body mass, and plasma
13  parameters and compared these with measurements from control birds prior to and after
14  manipulation (i.e. implantation of $T_3$ pellets or sham implantation). We predicted that
15  increased plasma $T_3$ levels, which will increase the energy expenditure of incubating female
16  eiders, should have immunosuppressive effects. Implanted females should therefore show
17  lower T-cell-mediated immune responses and/or a reduced immunoglobulin index than
18  control birds.
19
20
21
22
23
24
25

**MATERIALS AND METHODS**

The study was conducted in a common eider duck colony on Storholmen Island, Kongsfjorden, Svalbard Archipelago (78°55' N, 20°07' E) between June and July 2005. This breeding colony contained about 900 nests. Females laid between 1 and 6 eggs, but a clutch size of 4 to 5 eggs was most common (49.35 % and 24.62 %, respectively, N=845). Eider ducklings are precocial and are cared for by the female only. Incubation lasts between 24 and 26 days (Korschgen, 1977). All birds started their incubation between June 7 and June 13, the main laying period for the colony. Ducks that laid their eggs after this period were not considered in this study. Ambient temperatures in June and July ranged from 2 to 10°C.

Sampling protocol

Nests were checked at least every two days throughout the study period. This was done to determine initial clutch size but also to investigate the rate of egg predation and nest desertion. A clutch of eggs was considered complete when no additional egg was laid during a two-day period (Erikstad and Tveraa, 1995). Female eiders were caught on their nests using a bamboo pole with a nylon snare. Blood was collected from the brachial vein within three minutes of capture, stored in tubes containing EDTA (an anticoagulant agent) and kept on ice until further processing. In the laboratory blood was centrifuged at 10,000 rpm for five minutes at 4°C. Plasma samples and blood cells were separately stored at –20°C and plasma was subsequently used to measure immunoglobulin, $T_3$, and corticosterone levels. After blood sampling, body size was recorded (wing and tarsus lengths) and birds were weighed with a portable electronic balance (± 2 g).

Experimental groups and $T_3$ implantation

A total of 42 females with a mean clutch size of 4 to 5 eggs (4.17 ± 0.38 eggs; mean ± SD) were used in this study and captured on three occasions. To prevent nest desertion, we only caught birds that were already incubating for at least five days. Since both components of the acquired immunity decrease during whole incubation (Bourgeon et al., 2006a), we limited the present study to a relatively short time period, so that natural immunological changes occurring as the fasting is proceeding would not have confounding effects. Females were split into two experimental groups: birds with implanted $T_3$ pellets (N=21) and control females (N=21). Females from both groups were caught after about 10 days of incubation (9.95 ± 0.33 days, N=42; mean ± SE). When taking the first blood sample, we also recorded body size and body mass. Before release, half of the females were implanted, while the others underwent the same procedure without actual implantation. All females were recaught 6 to 8 days later (16.57 ± 0.42 days, N=42; mean ± SE), when a second blood sample was taken and the PHA skin test was conducted (see below). Birds were again weighed before their release. Finally, all birds were captured 24 hours later to read the PHA skin test. Only nests which did not suffer any predation were included in our study.

$T_3$ pellets (100 mg, 21 day release, T-261) were purchased from Innovative Research of America (Sarasota, Florida, USA). In preliminary trials, this dose was sufficient to induce a marked increase in plasma $T_3$ levels, while also causing an increased loss in body mass (Bourgeon et al., unpublished observation). $T_3$ pellets were implanted subcutaneously at the back side of the birds' neck. For this, we shaved the skin of the concerned area and disinfected using alcohol and betadine (iodine solution). A small incision, equal to the size of the pellet was made and the implant was positioned underneath the skin. The skin was closed with a single stitch, using surgical thread. The wound was cleaned with betadine and sprayed with an aluminium powder.

T-cell-mediated immune response: PHA skin test

To evaluate the T-cell-mediated immune response, 100 μl of 5 mg.ml$^{-1}$ PHA (Sigma L 8754) in phosphate-buffered saline (PBS) were injected intradermally in the right wing-web. It has been shown previously that birds suffer little physiological stress from PHA injection (Merino et al., 1999). The left wing-web was injected with an equal volume of PBS as a control. The thickness of each wing-web was measured with a micrometer calliper (three readings), just prior to and 24 h after injection. The cell-mediated immune response was calculated as the difference in wing-web swelling between the mitogen-injected and control site. Since the PHA skin test could only be conducted once with each animal (a second injection would lead to a secondary immune response), we injected all birds 6 to 8 days after manipulation.

Immunoglobulin levels: ELISA test

The amount of serum immunoglobulins in avian blood has been assessed using a sensitive ELISA method. Commercial antichicken conjugate antibodies were used as reported by Martínez et al. (2003). This method has been validated in six wild avian species. Although the method has not been validated for Anseriforms, we assumed a linear cross-reactivity, despite the fact that Anseriforms have an additional immunoglobulin isotype (IgY), which is not found in other birds (Parham, 1995). We used the measurements obtained from these tests as an immunoglobulin index.

The serum dilution in eiders was determined by coating ELISA plates with serial dilutions of serum (100 μl) in carbonate-bicarbonate buffer (0.1M, pH=9.6) to investigate the linear range of the sigmoid curve. Data obtained from trials using the serum dilution nearest to the centre of its linear range were selected. To be coated, ninety-six-well ELISA plates were filled with 100 μl of diluted serum samples from female common eiders (two samples per female, diluted to 1/32000 in carbonate-bicarbonate buffer). The plates were first incubated for 1 h at 37°C and then incubated overnight at 4°C. After washing the plates once with 200 μl of a solution of phosphate buffer saline and Tween (PBS-Tween), 100 μl of a solution containing 5% powdered milk in PBS was added. After a second incubation (1 h at 37°C) and a wash with PBS-Tween buffer, 100 μl of antichicken antibodies, diluted 1:250 (Sigma A 9046), were added and the plates were incubated for 2 h at 37°C. After three washes, the plates were finally filled with 100 μl of a solution (peroxide diluted in ABTS (2,2'-azino-bis-(3-ethylbenzthiazoline-6-sulphonic acid)) 1:1000). Following incubation (1 h at 37°C), the plates were read using a 405 nm wavelength filter (Awareness Technology, Inc., Palm City, FL, USA).

Assessment of the $T_3$ and corticosterone levels

$T_3$ and corticosterone concentrations were determined in plasma by radioimmunoassay (RIA) in our laboratory using a $^{125}I$ RIA double antibody kit from ICN Biomedicals (Costa Mesa, CA, USA). The RIA for $T_3$ and corticosterone had an intra-assay variability of 5.1 % (N=20 duplicates) and 7.1 % (N=10 duplicates), respectively. Inter-assay variability was 7.0 % (N=59 duplicates) for $T_3$ and 6.5 % (N=15 duplicates) for corticosterone.

Statistical analyses

Statistical analysis was conducted with SPSS 12.0.1 (SPSS Inc., Chicago, IL, USA). Values are means ± standard error (SE), unless otherwise indicated. Since all data were normally distributed (Kolgomorov-Smirnov test, P>0.05), we used parametric tests. Repeated measure ANOVA was used to test for treatment effects on $T_3$ levels, body mass, corticosterone levels, and the immunoglobulin index. One-way ANOVA was used to test for the effects of the treatment on the T-cell-mediated immune response. Linear regressions were used to assess the relationships between all measured parameters.

**RESULTS**

Table 1 provides details about the female eiders used in this study. There was no significant difference between the two groups for any of the parameters measured (see Table 1) before manipulation. We also ensured that females from both groups were at a comparable incubation stage, when they were caught to undergo the PHA skin test (Table 1).

Effects of $T_3$ on body mass and corticosterone levels:

As expected, implants induced a significant increase in $T_3$ levels (repeated measures ANOVA: effects of repetition: $F_{1,40}=186.38$, $P<0.0001$; effects of group: $F_{1,40}=124.29$, $P<0.0001$; interaction: $F_{1,40}=150.03$, $P<0.0001$). After 6 days, $T_3$ levels were 4 times higher in implanted females than in control birds (Table 1). Body mass decreased significantly in $T_3$ implanted females (15 %), while this decrease was less pronounced in control females (9 %; repeated measures ANOVA: effects of repetition: $F_{1,40}=524.05$, $P<0.0001$; effects of group: $F_{1,40}=0.05$, $P=0.82$; interaction: $F_{1,40}=23.70$, $P<0.0001$). Consequently, body mass of $T_3$ females after 6 days was significantly lower than that of control females. In fact, body mass loss per day was 40 % greater in $T_3$ females than in control females (Table 1). However, linear regression analysis showed no significant relationship between $T_3$ levels and body mass, neither before nor after the treatment (Table 2). Corticosterone levels were not significantly affected by $T_3$ implants (repeated measures ANOVA: effects of repetition: $F_{1,40}=27.12$, $P<0.0001$; effects of group: $F_{1,40}=0.01$, $P=0.93$; interaction: $F_{1,40}=0.003$, $P=0.96$). While plasma corticosterone levels after 6 days were increased by 60 %, this increase was similar for both groups (Table 1). Again, we did not find a significant relationship between $T_3$ levels and plasma corticosterone, neither before nor after the treatment (Table 2). Similarly, the relationship between corticosterone levels and body mass was not significant, neither before nor after the treatment (Table 2).

Effects of $T_3$ on immunoglobulin index and T-cell-mediated immune response:

Implanted females showed a similar T-cell-mediated immune response than control females (Table 1), indicating that $T_3$ implants had no significant effect on this parameter. Similarly, the relationship between $T_3$ levels after implantation and the T-cell-mediated immune response was not significant (Table 2). By contrast, the immunoglobulin index was significantly decreased in implanted females (18 %), while it actually increased by about 5 % in control females (repeated measures ANOVA: effects of repetition: $F_{1,40}=2.15$, $P=0.15$; effects of group: $F_{1,40}=0.05$, $P=0.83$; interaction: $F_{1,40}=8.17$, $P=0.007$; Fig. 1). Surprisingly, there was a significant positive relationship between $T_3$ levels and the immunoglobulin index only before the treatment (Table 2). This would indicate that high $T_3$ levels were associated with a high immunoglobulin index before the treatment but not thereafter (Fig. 2).

There was no significant relationship between the two components of the acquired immunity after hormone implantation (Table 2). Hence, the immunoglobulin index was independent of the T-cell-mediated immune response. Also, there was no significant relationship between the immunoglobulin index and plasma corticosterone levels or body mass, neither before nor after the treatment (Table 2). Furthermore, the T-cell-mediated immune response was not significantly related to neither plasma corticosterone levels, nor body mass (Table 2).

1  **DISCUSSION**

3  Previous studies have shown that the acquired immunity is significantly decreased during the
4  incubation fast of female common eiders (Hanssen et al., 2004; Bourgeon et al., 2006a). To
5  save energy, $T_3$ is adaptively decreased during fasting in most bird species (Harvey and
6  Klandorf, 1983). This raises the question of the relationship between thyroid hormones and
7  the immune system. Hence, the main objective of the current study was to investigate the
8  endocrine mechanisms underlying the immunosuppression reported during fasting in
9  incubating birds. To this end we assessed the effects of exogenous $T_3$ on both components of
10 the acquired immune system in free-ranging female eider ducks.

11         Experimentally increased plasma $T_3$ levels affected only one of the two components of
12 the acquired immunity. While the immunoglobulin index was significantly decreased in
13 implanted females, when compared with control females, the T-cell-mediated immune
14 response was not significantly affected by the treatment. Interestingly, while exogenous $T_3$
15 decreased the immunoglobulin index, high $T_3$ levels before the treatment were associated with
16 a high immunoglobulin index. Similar to previous results (Bourgeon et al., 2006a), we did not
17 find a significant relationship between both components of the acquired immunity. This
18 supports the view that variations in one component are not necessarily a reliable indicator of
19 changes in the other, as suggested by Norris and Evans (2000). In the present study, the
20 immunoglobulin index was more sensitive to $T_3$ treatment than was the T-cell-mediated
21 immune response. This is in accordance with previous results (Bourgeon et al., 2006a), which
22 showed that, for the same time period, the immunoglobulin index decreases two times faster
23 than the T-cell-mediated immune response in incubating eiders. We can therefore not exclude
24 the possibility that effects of $T_3$ on the T-cell-mediated immune response might require more
25 time and/or higher $T_3$ concentrations.

1    The treatment did not induce any significant changes in corticosterone levels. After 6
2    days, corticosterone levels in $T_3$ implanted females were similar to that of control females.
3    Glucocorticoids, which are secreted during stressful activities, are an essential component of
4    the endogenous immunoregulatory network (Apanius, 1998). Furthermore, Råberg et al.
5    (1998) suggested that corticosterone reduces the adaptive immune function. However, since
6    corticosterone levels in our study were not affected by the treatment, the observed decrease in
7    the immunoglobulin index might be mediated by $T_3$, independently of corticosterone. Indeed,
8    an experimental study performed on the same species showed that exogenous corticosterone
9    significantly decreased immunoglobulin index (Bourgeon and Raclot, 2006). Moreover, in the
10   current study, we did not find a significant relationship between corticosterone levels and the
11   immunoglobulin index or the T-cell-mediated immune response, neither before, nor after the
12   treatment. This is in agreement with results from an earlier study (Bourgeon et al., 2006a),
13   which found that plasma corticosterone levels did not vary throughout the incubation period
14   of female eiders.

15   $T_3$ partially regulates thermoregulation and energy metabolism in mammals and birds
16   (McNabb, 1995). Accordingly, $T_3$ implanted females in our study lost significantly more
17   weight than did control females. Norris and Evans (2000) proposed that immunocompetence
18   may be fixed by the BMR, which is determined by energy reserves. Such a view on the
19   relationship between energy metabolism and immunocompetence is supported by the
20   existence of nutritional and endocrine factors that regulate both of these processes (Apanius,
21   1998). Since eider ducks do not feed during incubation, limited resources might be allocated
22   for the reconstitution of tissues affected by gluconeogenesis, and, therefore be unavailable for
23   the maintenance of immunity (Saino et al., 2002). Moreover, adipose tissue is no longer
24   regarded only as a fat store but also as an important endocrine organ, responsible for the
25   synthesis and secretion of several hormones and proteins (Ahima and Flier, 2000). It has recently been described as an active participant in the regulation of essential and prominent body processes, such as immune homeostasis (Matarese and La Cava, 2004). Some adipose humoral signals are generated in proportion to fat stores and act on feedback control systems to influence food intake and energy expenditure (Matarese and La Cava, 2004). This latter fact raises the question of the hormonal control of the immune system by body reserves (Demas and Sakaria, 2005). The peptide hormone leptin is secreted primarily by adipose tissue and has been shown to enhance a variety of immunological parameters in mammals (Lord et al., 1998; Faggioni et al., 2001) and birds (Lõhmus et al., 2004). Since circulating levels of leptin are generally proportional to the amount of body fat (Lõhmus and Sundström, 2004; Matarese et al., 2005), decreases in body fat stores may affect immunity via changes in endocrine signalling (Demas and Sakaria, 2005). Exogenous administration of $T_3$ is likely to increase BMR, further depleting body energy stores. Consequently, leptin levels might be lowered in $T_3$ implanted birds, when compared with the controls. While the effects induced by leptin might favour survival under hostile conditions, concomitant starvation leads to immunosuppression and impaired fertility, because energy-consuming processes are switched off by leptin (Matarese and La Cava, 2004). The immunosuppressive effect of exogenous $T_3$ could therefore be mediated by leptin. For an examination of the role of this hormone in the immune function of fasting eiders, direct manipulation of leptin concentrations would be helpful (Lõhmus et al., 2004). In addition, since we found that an experimentally induced increase in the utilization of body reserves led to immunosuppression, it seems likely that immunocompetence is related to the amount of body fat available. Initial body condition and the subsequent utilization of endogenous reserves might therefore determine the observed variations in immune functioning throughout incubation. Females with a good initial body condition would be less prone to suffer immunosuppression during incubation than females with a bad initial body condition.

While the main goal of our study was to investigate the physiological mechanisms underlying immunosuppression in breeding eiders, it might also provide evidence regarding its adaptive significance. Two hypotheses, which are not mutually exclusive, have been proposed to explain the immunosuppression observed during breeding. The immunopathology-avoidance hypothesis states that during heavy physical workloads, such as encountered during reproduction, the risk of an autoimmune response increases (Råberg et al., 1998). To decrease the risk of auto immunopathology, immunocompetence is down regulated. Råberg et al. (1998) suggested that such a down regulation would be mediated by corticosterone. However, in the present study we did not find a significant relationship between corticosterone levels and the T-cell-mediated immune response or the immunoglobulin index. Nevertheless, our present results do not allow us to exclude this hypothesis. Notably, we did not measure the heat shock protein (HSP) levels which are more appropriate for detecting chronic or long-term exposure to stressors (Martinez-Padilla et al., 2004 ; Tomás et al., 2004). HSPs have been shown to significantly increase over incubation in breeding eiders, further partially supporting the immunopathology-avoidance hypothesis in this species (Bourgeon et al., 2006b). The second hypothesis, termed the resource-limitation hypothesis, predicts that the investment in costly behaviours, such as reproduction, will reduce the amount of resources available to other systems, such as the immune system (Råberg et al., 1998). For this hypothesis to be valid, an energetic or nutritional cost associated with the maintenance and activation of the immune system is required (Råberg et al., 1998). While evidence for an energetically costly immune response is still equivocal (Råberg et al., 1998; Eraud et al., 2005; Verhulst et al., 2005), the present study shows that an increase in energy expenditure, caused by $T_3$ administration, had a negative effect on the females' immunoglobulin index. This would lend support to the latter hypothesis, illustrating the trade-off between one component of immunity and other resource demanding activities.

In conclusion, exogenous $T_3$ decreased only one component of the acquired immune function in incubating female eiders. While their immunoglobulin index was significantly decreased after $T_3$ administration, their T-cell-mediated immune response was not affected. Since $T_3$ implants did not induce changes in corticosterone concentrations, this would suggest that glucocorticoids were not involved in the observed decrease in the immunoglobulin index. Weight loss was significantly greater in $T_3$ implanted females than in control females. The immunosuppressive effect of $T_3$ might therefore be mediated by its effects on energy expenditure and/or body fat reserves. In fact, leptin, which conveys information on energy availability, could be responsible for the observed immunosuppression in our study. Further experiments are required to shed more light onto the relationship between leptin, body condition and the immune system in incubating female eider ducks. Finally, despite the fact that short-term energetic costs have not been observed in female eiders injected with three different non-pathogenic antigens while their survival was compromised (Hanssen et al., 2004), the adaptive significance of immunosuppression during incubation in eider ducks still remains to be documented.

**ACKNOWLEDGMENTS**

Financial support for this work was provided by the Institut polaire français Paul Emile Victor (IPEV). S. Bourgeon was the recipient of a fellowship from the French MENRT during this study. This study was supported by the Norwegian Animal Research Authority and the Ethic Committee of the Institut Polaire Français Paul-Emile Victor. Permission to work on Common Eiders was granted by the Governor of Svalbard.
Bourgeon and Raclot GCE-06-68 Revised manuscript                    16

6
7
8
9
10
11
12
13
14
15
16
17
18
19
20
21
22
23
24
25

**TABLES**

**Table 1.** Profiles for both experimental groups of free-ranging incubating female eiders, before and after implantation of $T_3$ pellets. Control animals were sham implanted. Values are means ± SE.

| Before implantation | Group 1: $T_3$ females (N=21) | Group 2: Control females (N=21) | T-test | P |
|---|---|---|---|---|
| **Initial clutch size (eggs)** | 4.19 ± 0.09 | 4.14 ± 0.08 | 0.40 | 0.69 |
| **Tarsus length (cm)** | 6.15 ± 0.03 | 6.08 ± 0.04 | 1.57 | 0.12 |
| **Wing length (cm)** | 29.30 ± 0.11 | 29.00 ± 0.16 | 1.56 | 0.12 |
| **Incubation stage at sampling (days)** | 10.09 ± 0.49 | 9.81 ± 0.45 | 0.43 | 0.67 |
| **Body mass at sampling (g)** | 1836 ± 31 | 1784 ± 28 | 1.25 | 0.13 |
| **Immunoglulin index (absorbance units)** | 0.85 ± 0.04 | 0.78 ± 0.04 | 1.51 | 0.25 |
| **Triiodothyronine (pg.ml$^{-1}$)** | 12.26 ± 0.74 | 10.14 ± 0.89 | 1.84 | 0.07 |
| **Corticosterone (ng.ml$^{-1}$)** | 12.17 ± 1.30 | 11.94 ± 1.01 | 0.14 | 0.89 |

| After implantation | Group 1: $T_3$ females (N=21) | Group 2: Control females (N=21) | T-test | P |
|---|---|---|---|---|
| **Incubation stage at sampling (days)** | 17.14 ± 0.60 | 16.00 ± 0.56 | 1.38 | 0.17 |
| **Body mass at sampling (g)** | 1590 ± 29 | 1624 ± 27 | -0.87 | 0.39 |
| **Body mass loss per day (g)** | 35.62 ± 1.72 | 25.62 ± 1.44 | 4.46 | <0.0001 |
| **Immunoglulin index (absorbance units)** | 0.72 ± 0.04 | 0.82 ± 0.04 | -1.63 | 0.11 |
| **T-cell-mediated immune response (mm)** | 0.98 ± 0.11 | 0.92 ± 0.11 | 0.42 | 0.68 |
| **Triiodothyronine (pg.ml$^{-1}$)** | 44.76 ± 2.25 | 11.90 ± 1.28 | 12.68 | <0.0001 |
| **Corticosterone (ng.ml$^{-1}$)** | 19.45 ± 1.84 | 19.38 ± 1.82 | 0.03 | 0.98 |

**Table 2.** Results for linear regressions between immune parameters, body mass, $T_3$, and corticosterone levels in free-ranging incubating female eiders.

| Before implantation | Body mass (g) | Immunoglulin index (absorbance units) | Triiodothyronine (pg.ml$^{-1}$) | Corticosterone (ng.ml$^{-1}$) |
|---|---|---|---|---|
| **Body mass (g)** | - | $F_{1,41}$=2.16, P=0.15 | $F_{1,41}$=2.90, P=0.10 | $F_{1,41}$=0.69, P=0.41 |
| **Immunoglulin index (absorbance units)** | | - | $F_{1,41}$=9.91, P=0.003 | $F_{1,41}$=0.52, P=0.47 |
| **Triiodothyronine (pg.ml$^{-1}$)** | | | - | $F_{1,41}$=0.65, P=0.43 |

| After implantation | Body mass (g) | Immunoglulin index (absorbance units) | T-cell-mediated immune response (mm) | Triiodothyronine (pg.ml$^{-1}$) | Corticosterone (ng.ml$^{-1}$) |
|---|---|---|---|---|---|
| **Body mass (g)** | - | $F_{1,41}$=0.01, P=0.91 | $F_{1,41}$=0.20, P=0.66 | $F_{1,41}$=0.35, P=0.56 | $F_{1,41}$=0.05, P=0.82 |
| **Immunoglulin index (absorbance units)** | | - | $F_{1,41}$=0.35, P=0.56 | $F_{1,41}$=1.03, P=0.31 | $F_{1,41}$=1.11, P=0.30 |
| **T-cell-mediated immune response (mm)** | | | - | $F_{1,41}$=0.38, P=0.54 | $F_{1,41}$=0.72, P=0.40 |
| **Triiodothyronine (pg.ml$^{-1}$)** | | | | - | $F_{1,41}$=0.01, P=0.92 |

**FIGURE LEGENDS**

**Figure 1.** Effects of $T_3$ administration on the immunoglobulin index in free-ranging incubating female eiders. Shown is the immunoglobulin index of $T_3$ implanted () and control females () before and after manipulation. Values are mean ± SE. $T_3$, triiodothyronine. Lower case a and b indicate a significant difference between groups (T-tests).

**Figure 2.** Relationship between $T_3$ level and the immunoglobulin index in free-ranging incubating female eiders before implantation. $T_3$, triiodothyronine.

1 **FIGURE 1.**

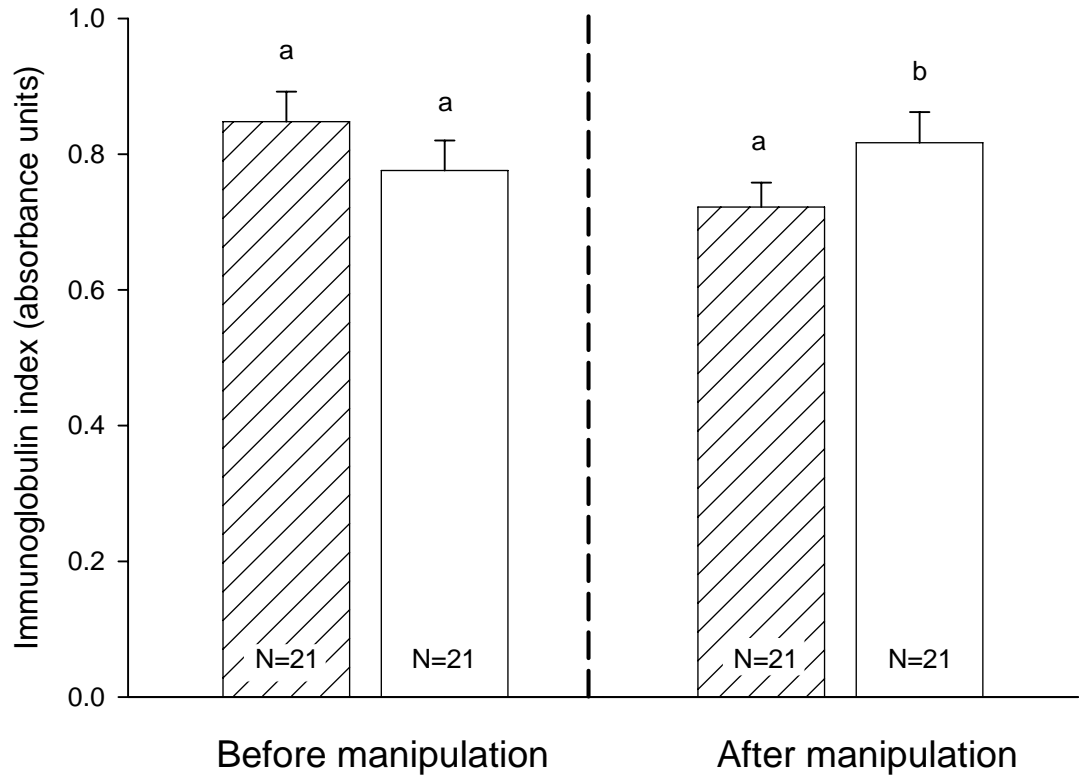

Before manipulation      After manipulation

1 **FIGURE 2.**

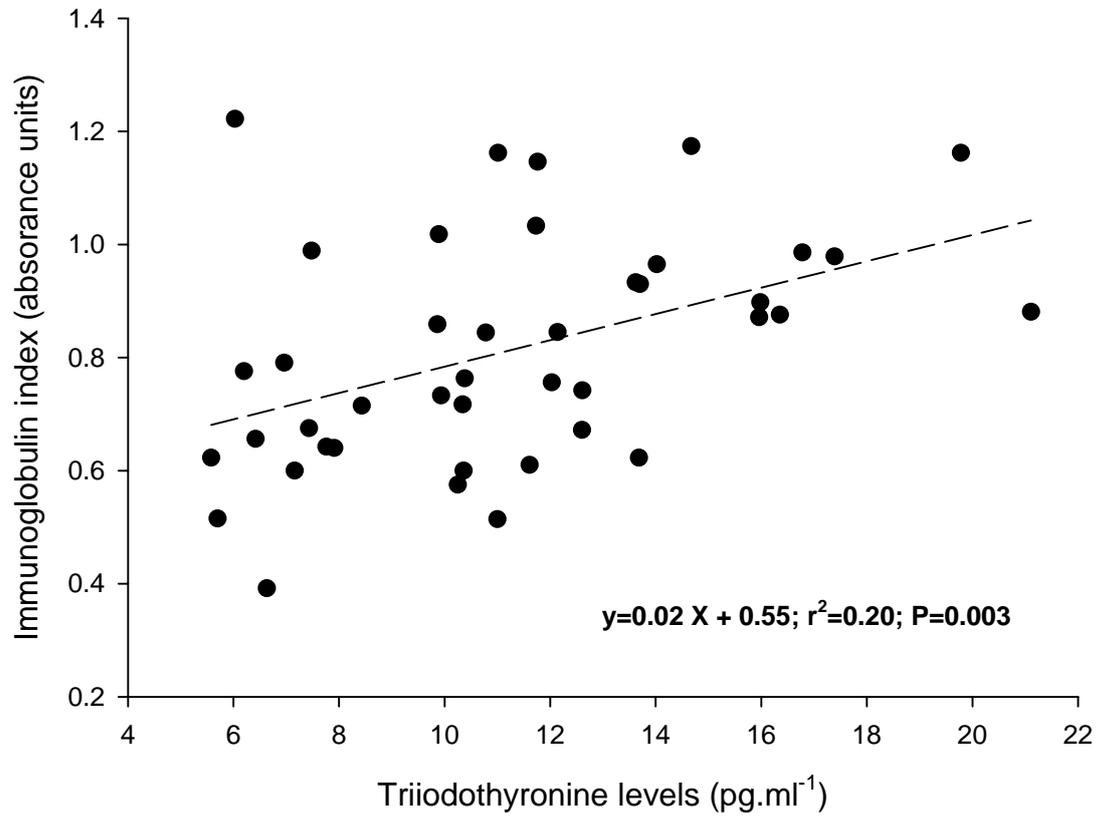